\documentclass[prl,aps,twocolumn,floatfix,longbibliography]{revtex4-2}

\usepackage{amsmath,amssymb}
\usepackage{bm}
\usepackage{graphicx}
\usepackage{hyperref}

\begin{document}

\title{Anomalous radiation pressure in strong-field ionization driven by quantum light}

\author{Xiaodan Mao}
\affiliation{State Key Laboratory of Dark Matter Physics, Key Laboratory for Laser Plasmas (Ministry of Education) and School of Physics and Astronomy, Shanghai Jiao Tong University, Shanghai 200240, China}

\author{Feng He}
\affiliation{State Key Laboratory of Dark Matter Physics, Key Laboratory for Laser Plasmas (Ministry of Education) and School of Physics and Astronomy, Shanghai Jiao Tong University, Shanghai 200240, China}

\author{Pei-Lun He}
\email{peilunhe@sjtu.edu.cn}
\affiliation{State Key Laboratory of Dark Matter Physics, Key Laboratory for Laser Plasmas (Ministry of Education) and School of Physics and Astronomy, Shanghai Jiao Tong University, Shanghai 200240, China}

\date{\today}
\begin{abstract}
We show that in strong-field ionization driven by bright squeezed vacuum, the mean longitudinal photoelectron momentum scales with the mean incident intensity $I$ as $I^{2/3}$, rather than linearly as under coherent-light driving.
This anomalous scaling originates from a field-amplitude saddle-point structure: nonlinear tunneling selects two dominant field amplitudes of equal magnitude and opposite sign from the broad quantum fluctuations.
Tracing over the final photon state erases their relative-phase information, leaving their common magnitude to determine the field-dependent longitudinal momentum shift and its $I^{2/3}$ scaling.
Photon-number resolution instead preserves this coherence, producing parity-dependent modulations of the longitudinal momentum transfer through time-domain double-slit interference between two field pathways separated by half a cycle.
These results establish longitudinal momentum transfer as a distinct observable in strong-field quantum optics that encodes both the photon statistics and the field coherence of intense quantum light.
\end{abstract}

\maketitle

Radiation pressure is a direct manifestation of momentum exchange between light and matter. For classical light, it scales linearly with the optical intensity, reflecting the electromagnetic momentum flux carried by the Poynting vector~\cite{JacksonElectrodynamics}. Microscopically, each exchange of an energy quantum $\hbar\omega$ is accompanied by a momentum transfer $\hbar k$, with $k=\omega/c$~\cite{Einstein1905Photoelectric,Compton1923Scattering}. 
In photoionization, part of this momentum is carried away by the emitted electron, an effect anticipated by early theory but confirmed experimentally only nine decades later~\cite{Sommerfeld1930Photoeffect,Schmidt2024BindingEnergyNondipole}.

In the strong-field regime, this momentum exchange manifests itself as a radiation-pressure-induced shift of the photoelectron distribution along the propagation direction~\cite{Smeenk2011PhotonMomentum,Reiss2013RelativisticEffects,Chelkowski2014MomentumSharing,He2017BeyondDipole}. Because this shift is absent in the dipole approximation, the longitudinal photoelectron momentum provides a sensitive probe of nondipole dynamics~\cite{Klaiber2013UnderBarrier,Ludwig2014DipoleBreakdown,Hartung2021Electric,Lin2022EnergyPeaks}. Across single-photon and multiphoton ionization regimes, both the magnitude of the transferred momentum and its partition between the emitted electron and the residual ion have been established~\cite{Cricchio2015MomentumPartition,Grundmann2020PhotoionBackward,He2023DoubleSlitIonDynamics,Mao2025MomentumPartition}. Nondipole signatures have also been identified in above-threshold ionization~\cite{Hartung2019Magnetic,Willenberg2019Subcycle,Haram2019RelativisticNondipole,Lin2022HighOrderATI}, high-harmonic generation~\cite{Walser2000HHGNondipole,Klaiber2017LimitsRescattering}, ionization-time-delay observables~\cite{Grundmann2020Zeptosecond,Liang2021NondipoleTwoElectronInterference,He2022NondipoleTimeDelay,Liang2024AttosecondNondipole}, and nonsequential double ionization~\cite{Lin2022MagneticCorrelation,Dou2026NondipoleCorrelationXe}. 
In these studies, however, the driving field is coherent or classically prescribed, and the associated radiation-pressure momentum transfer follows the conventional linear scaling with the laser intensity $I$.

Phase-space formulations of interactions with intense quantum light represent the field as an ensemble of coherent-state components, each associated with a classical driving field~\cite{Gorlach2023QuantumHHG,Fang2023QuantumIonization,Lyu2025PhotonQuantumStatisticsATI,zhou2026attosecond}. Such probability-level averages, however, discard the phase coherence between field components, whereas photon-number-resolved observables can retain this coherence~\cite{Mao2025Benchmarking}. Bright squeezed vacuum (BSV) is particularly suited to the bright-field regime because it combines a macroscopic photon population with exceptionally broad photon-number and quadrature fluctuations~\cite{Iskhakov2012Superbunched,Spasibko2017Multiphoton,lyu2026generation}. These broad fluctuations make nonlinear strong-field observables particularly sensitive to photon statistics, even at low mean intensity. Quantum-statistical effects have been identified in tunneling and above-threshold ionization~\cite{WangLai2023QuantumATI,yrpf-r7p6,Habibovic2025IntensityDependentQuantumLight}, continuum-electron dynamics~\cite{EvenTzur2023PhotonStatisticsForce,EvenTzurCohen2024BSVMotion,Heimerl2025QuantumNeedle}, high-harmonic generation~\cite{Theidel2024QuantumOpticalHHG,Rasputnyi2024BSVHHG,Yi2025MassivelyEntangledHHG,Liu2026TheoryQuantumHHG}, relativistic strong-field QED processes~\cite{Khalaf2023ComptonQuantumLight,DiPiazza2026NonlinearComptonSqueezed,Ge2026NonlinearBreitWheelerSqueezed}, and molecular dissociation~\cite{Long2025QuantumDissociation}. 
Despite this progress, whether quantum statistics can modify the propagation-direction momentum transfer in atomic ionization, the microscopic signature of radiation pressure, remains unresolved. The recent demonstration of BSV-driven tunneling ionization~\cite{Jiang2026QuantumTunnelling} makes this question experimentally timely.

Here we show that BSV-driven strong-field ionization exhibits an anomalous radiation-pressure law, $\langle p_z\rangle\propto I^{2/3}$, rather than the linear scaling obtained under coherent-light driving. A joint light-front saddle-point analysis reveals an underlying coherent two-branch field structure formed by two equal-magnitude, opposite-sign field configurations selected from the broad BSV fluctuations. 
Tracing over the final photon state erases the relative-phase information between these branches, leaving only their common magnitude to determine the anomalous longitudinal momentum shift.
Photon-number resolution instead retains their coherent superposition and exposes the underlying time-domain double-slit interference between two half-cycle-separated field pathways, producing parity-dependent modulations of the longitudinal momentum transfer and ionization yield.

In the bright-field limit, tracing over the final photon state reduces the BSV-driven photoelectron spectrum to a Gaussian phase-space average of coherent-field ionization probabilities, a reduction benchmarked against fully quantized calculations~\cite{Mao2025Benchmarking},
\begin{equation}
P(\mathbf p)=
\int_{-\infty}^{\infty}
\frac{dA_\alpha}{\sqrt{2\pi}A_{\mathrm{rms}}}
\exp\left(-\frac{A_\alpha^2}{2A_{\mathrm{rms}}^2}\right)
\left|M_{A_\alpha}(\mathbf p)\right|^2 .
\label{eq:field-average-main}
\end{equation}
Here, $A_\alpha$ denotes the vector-potential amplitude of a coherent-state component, while $A_{\mathrm{rms}}$ characterizes the width of the BSV amplitude distribution. We define $A_{\mathrm{rms}}$ through the mean incident intensity,
\[
I=2\omega^2A_{\mathrm{rms}}^2,
\qquad
A_{\mathrm{rms}}^2
\equiv
\frac{2\epsilon_V^2}{1-\tanh r}
\approx
\epsilon_V^2e^{2r},
\]
where the last relation holds in the bright-BSV limit. For the circular polarization considered here, each coherent-state component is represented by
\begin{equation}
\mathbf A(\eta;A_\alpha)
=
A_\alpha f(\eta)
\left[
\sin\phi(\eta)\,\mathbf e_x
-\cos\phi(\eta)\,\mathbf e_y
\right].
\label{eq:field-realization-main}
\end{equation}
Here, $\phi(\eta)=\omega\eta-\theta/2$, with $\theta$ the squeezing phase. The function $f(\eta)$ describes the Gaussian temporal envelope of the collective wave packet formed by the frequency components of the experimentally generated BSV pulse~\cite{Wunsche2004QuantizationBeams,Jiang2026QuantumTunnelling}.
We use atomic units unless stated otherwise.

For each field realization entering Eq.~\eqref{eq:field-average-main}, the ionization amplitude is evaluated within the leading-order light-front nondipole strong-field approximation~\cite{He2022Subbarrier},
\begin{equation}
M_{A_\alpha}(\mathbf p)
=
-i\int d\eta\,
\mathcal D_{A_\alpha}(\mathbf p,\eta)
e^{-iS_{A_\alpha}(\mathbf p,\eta)},
\label{eq:coherent-amplitude-main}
\end{equation}
where $\mathcal D_{A_\alpha}(\mathbf p,\eta)$ denotes the bound--continuum transition prefactor for the corresponding field realization, and the nondipole semiclassical action is given by
\begin{equation}
S_{A_\alpha}(\mathbf p,\eta)
=
\int_{\eta}^{\infty}d\tau\,
\frac{
\left[\widetilde{\mathbf p}
+\mathbf A(\tau;A_\alpha)\right]^2+\kappa^2
}{2\Lambda}.
\end{equation}
Here, $\eta=t-z/c$ is the light-front time~\cite{Dirac1949Forms}, and $\widetilde{\mathbf p}=(p_x,p_y,\widetilde p_z)$ is the light-front momentum, with $\widetilde p_z=p_z-(\mathbf p^2/2+I_p)/c$. We further define $\Lambda=1-\widetilde p_z/c$ and $\kappa=\sqrt{2I_p}$, where $I_p$ is the ionization potential. 
For circularly polarized fields, in which rescattering is strongly suppressed, SFA-based descriptions have been shown to capture the main features of direct-ionization momentum distributions and the nonadiabatic trends obtained from time-dependent Schr\"odinger equation calculations~\cite{Barth2014NumericalVerification}.
The derivation and numerical details are provided in the Supplemental Material~\cite{supp}.

\begin{figure}[t]
\centering
\includegraphics[width=\columnwidth]{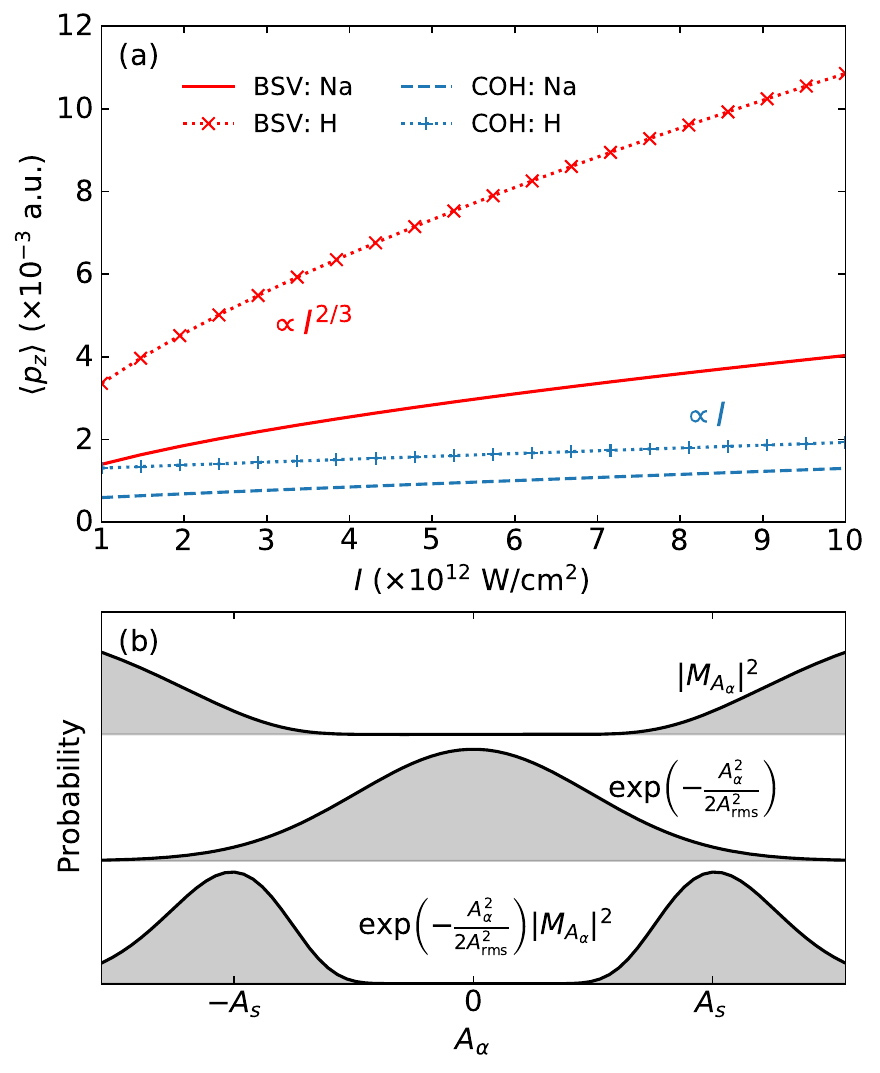}
\caption{
Anomalous radiation pressure and field-amplitude selection.
(a) Mean longitudinal photoelectron momentum versus the mean incident intensity for H and Na, showing the linear coherent-state response and the $I^{2/3}$ scaling for BSV in the tunneling regime.
The laser frequency is $\omega=0.0288$ a.u. ($1580\,\mathrm{nm}$) and the pulse duration is $L=10T$, with $T=2\pi/\omega$.
(b) Schematic of the field-amplitude selection mechanism.
The BSV-weighted ionization probability develops two maxima at the dominant field-amplitude saddles $\pm A_s$.
The curves are vertically offset for clarity.
}
\label{fig:pzshift}
\end{figure}

Figure~\ref{fig:pzshift}(a) contrasts the intensity dependence of the longitudinal momentum shift under coherent-state and BSV driving. The coherent-state result shows the conventional linear dependence on $I$, whereas the BSV result approaches the anomalous scaling $\langle p_z\rangle\propto I^{2/3}$ in the tunneling regime.
To uncover the origin of this anomalous scaling, we perform a joint steepest-descent analysis of the field integral in Eq.~\eqref{eq:field-average-main} and the two ionization-time integrals in $M_{A_\alpha}(\mathbf p)M_{A_\alpha}^*(\mathbf p)$~\cite{supp}. The stationary field amplitude and complex ionization times satisfy
\begin{equation}
\begin{aligned}
&\left[\widetilde{\mathbf p}+\mathbf A(\eta_{j,s};A_s)\right]^2+\kappa^2=0,
\qquad j=1,2,\\
&\frac{A_s}{A_{\mathrm{rms}}^2}
=i\left[
\partial_{A_\alpha}S_{A_\alpha}(\mathbf p,\eta_{2,s})
-\partial_{A_\alpha}S_{A_\alpha}(\mathbf p,\eta_{1,s})
\right]_{A_\alpha=A_s}.
\end{aligned}
\label{eq:photon-traced-saddles-main}
\end{equation}
The first line determines the two ionization-time saddles. Unlike conventional strong-field saddle-point treatments with a prescribed field amplitude, the second line promotes the fluctuating amplitude itself to a saddle variable whose value is selected by the balance between the BSV amplitude weight and the strongly field-dependent tunneling probability, as illustrated schematically in Fig.~\ref{fig:pzshift}(b). For the parameters considered, the dominant solutions lie in the self-conjugate sector, with $\eta_{2,s}=\eta_{1,s}^*$ and $A_s\in\mathbb R$. At a given saddle value $A_s$, the electron dynamics is governed by the same coherent-field ionization amplitude $M_{A_s}(\mathbf p)$ that describes ionization by a classically prescribed field of amplitude $A_s$. The full saddle structure, including the complex branches and the derivation of the adiabatic limit used below, is presented in the End Matter.

For circular polarization and vanishing transverse exit velocity, the leading-order adiabatic solution consists of two branches with opposite field amplitudes,
\begin{equation}
A_s\approx\pm\sqrt{2I_p}
\left(\frac{4A_{\mathrm{rms}}^2}{3\omega}\right)^{1/3},
\qquad
A_s^2\propto I^{2/3}.
\label{eq:radiation-pressure-main}
\end{equation}
These branches correspond to the two field-amplitude saddles represented schematically by the maxima at $\pm A_s$ in Fig.~\ref{fig:pzshift}(b). For photon-traced observables, the common magnitude of the two branches sets the field-dependent radiation-pressure scale, approximately $A_s^2/(2c)$. Substituting $A_{\mathrm{rms}}^2=I/(2\omega^2)$ then yields
$\frac{A_s^2}{2c}
=
\frac{I_p}{\omega^2c}
\left(\frac{2}{3}\right)^{2/3}
I^{2/3}$.
For coherent-state driving, by contrast, the prescribed field amplitude yields the conventional field-dependent contribution $I/(4\omega^2c)$. Including the nearly field-independent subbarrier offset $I_p/(3c)$~\cite{Hartung2019Magnetic}, the mean longitudinal momenta become
\begin{equation}
\begin{aligned}
\langle p_z\rangle_{\mathrm{COH}}
&\simeq
\frac{I}{4\omega^2c}
+\frac{I_p}{3c},\\
\langle p_z\rangle_{\mathrm{BSV}}
&\simeq
\frac{I_p}{\omega^2c}
\left(\frac{2}{3}\right)^{2/3}
I^{2/3}
+\frac{I_p}{3c}.
\end{aligned}
\label{eq:momentum-scaling-main}
\end{equation}
Under BSV driving, the conventional linear radiation-pressure law is thus replaced by the anomalous $I^{2/3}$ scaling. In this adiabatic limit, the slope of the field-dependent contribution for coherent light is independent of $I_p$, and the target dependence enters only through the subbarrier offset. For BSV, by contrast, $I_p$ also determines the prefactor of the field-dependent contribution. Corrections beyond the leading-order adiabatic approximation modify the longitudinal momentum shift quantitatively but preserve the characteristic $I^{2/3}$ scaling. Further robustness tests, addressing nonadiabatic effects, continuum Coulomb effects, and over-the-barrier corrections, are presented in the Supplemental Material~\cite{supp}.

\begin{figure}[t]
\centering
\includegraphics[width=\columnwidth]{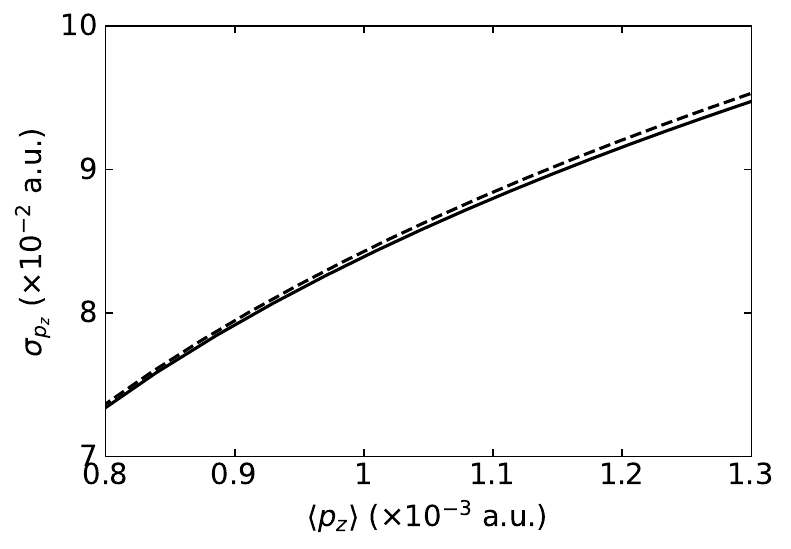}
\caption{Longitudinal momentum width versus the mean propagation-direction momentum for Na. The BSV (solid) and coherent-state (dashed) results nearly overlap. The remaining parameters are the same as in Fig.~\ref{fig:pzshift}.}
\label{fig:scaling}
\end{figure}

Before turning to photon-number resolution, we test the effective-field picture through the relation between the mean longitudinal momentum shift $\langle p_z\rangle$ and the width $\sigma_{p_z}$ of the longitudinal momentum distribution. Figure~\ref{fig:scaling} plots $\sigma_{p_z}$ directly against $\langle p_z\rangle$, thereby eliminating the incident intensity as an explicit parameter. Despite their markedly different dependences on the incident intensity, the BSV and coherent-state results nearly collapse onto a common curve. The relation between longitudinal momentum broadening and momentum transfer is therefore nearly insensitive to the photon statistics of the driving field. Within the effective-field picture, this near overlap indicates that quantum statistics primarily determines which field amplitudes are sampled during nonlinear ionization. The small residual deviations arise from subleading saddle corrections, as discussed in the End Matter.

The photon-traced analysis above captures only the phase-insensitive remnant of the two-branch field structure. To uncover the coherent origin of this structure, we now resolve the final photon number. This requires the coherent-state amplitudes to be combined before taking the modulus squared, allowing coherence between the two branches to remain observable. Within the coherent-state path-integral formulation benchmarked against fully quantized calculations~\cite{Mao2025Benchmarking}, projection onto the final photon number gives
\begin{equation}
P_n(\mathbf p)=\left|\int\frac{d^2\alpha}{\pi}\,
\langle n|\alpha\rangle\langle\alpha|\xi\rangle
M_\alpha(\mathbf p)\right|^2,
\label{eq:joint-main}
\end{equation}
where $|\xi\rangle$ denotes the BSV state, with $\xi=re^{i\theta}$.
When the final photon number is unresolved, summing over $n$ invokes the completeness relation
$\sum_n\langle\alpha'|n\rangle\langle n|\alpha\rangle
=\langle\alpha'|\alpha\rangle$.
In the macroscopic-field limit, this overlap becomes sharply localized on the diagonal of coherent-state phase space, reducing the electron spectrum to the diagonal phase-space average in Eq.~\eqref{eq:field-average-main}~\cite{EvenTzur2023PhotonStatisticsForce}. 
As a result, photon tracing removes the relative-phase information between the two branches, whereas projection onto a definite photon number retains it. We now use Eq.~\eqref{eq:joint-main} to analyze the photon-number-conditioned momentum transfer along the propagation direction.

The photon-number projection modifies the field-saddle structure, as detailed in the End Matter. After analytic continuation, with $\alpha$ and $\bar\alpha$ treated as independent variables, the saddle equations admit two dominant solutions,
\begin{equation}
\alpha_s^{(\pm)}
\simeq
\pm i\sqrt{n}\,e^{i\theta/2},
\qquad
\bar\alpha_s^{(\pm)}
\simeq
\mp i\sqrt{n}\,e^{-i\theta/2}.
\label{eq:fock-saddles-main}
\end{equation}
These solutions reveal the coherent amplitude-level structure underlying the opposite-sign field saddles found in the photon-traced analysis. They correspond to two field configurations with equal magnitudes and a relative phase of $\pi$, and the associated ionization amplitudes interfere coherently in Eq.~\eqref{eq:joint-main}. Their common vector-potential magnitude is
$A_n=2\epsilon_V\sqrt{n}$, where
$\epsilon_V=\sqrt{2\pi/(V\omega)}$ and $V$ is the effective pulse-mode volume~\cite{Wunsche2004QuantizationBeams}. 

\begin{figure}[t]
\centering
\includegraphics[width=\columnwidth]{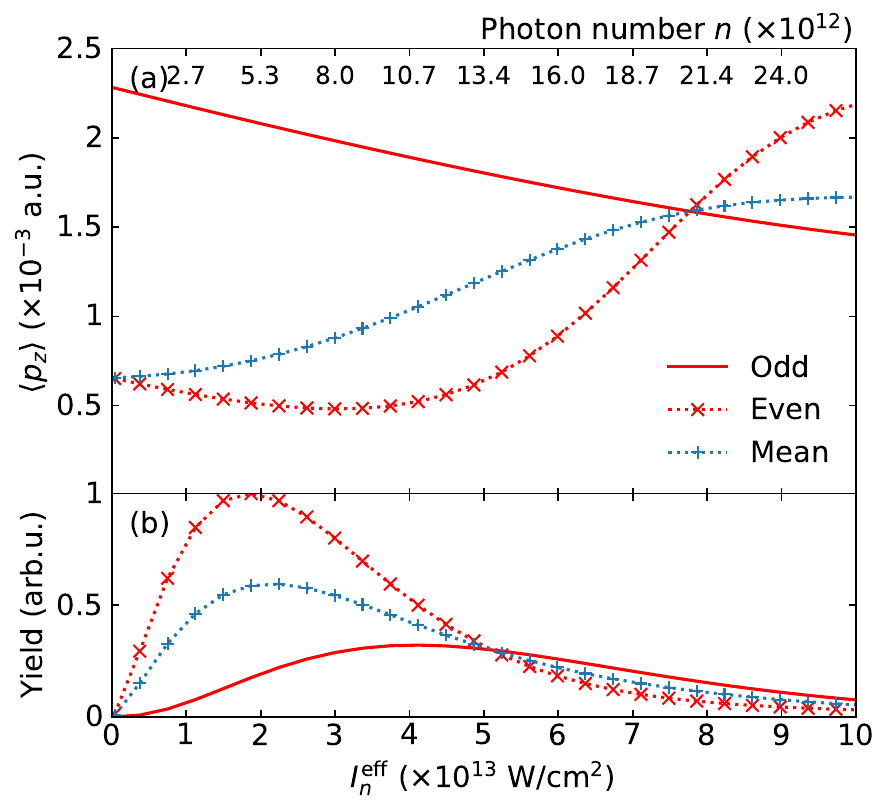}
\caption{Photon-number-resolved anomalous radiation pressure for Na. 
(a) Mean longitudinal momentum $\langle p_z\rangle_n$ resolved by photon-number parity, versus the effective intensity $I_n^{\mathrm{eff}}$ associated with the final $n$-photon channel. (b) Corresponding photon-number-resolved ionization probability. 
The laser wavelength is $400\,\mathrm{nm}$, with pulse duration $L=10T$, $\epsilon_V=3.2\times10^{-8}$~a.u., and mean intensity $I=1\times10^{13}\,\mathrm{W/cm^2}$.}
\label{fig:photon_resolved}
\end{figure}

Figure~\ref{fig:photon_resolved} shows the longitudinal momentum and ionization probability resolved by final photon number as functions of the effective intensity $I_n^{\mathrm{eff}}\equiv2\omega^2A_n^2$ associated with each $n$-photon channel.
A shorter wavelength of $400\,\mathrm{nm}$ is used here to resolve the parity-dependent interference structure more clearly.
Although the initial BSV contains only even Fock components, ionization populates both even and odd final channels through electron--field entanglement.
The overall channel dependence follows the associated field scale, with a pronounced even--odd alternation superimposed on this trend. At lower $I_n^{\mathrm{eff}}$, even channels are more probable and exhibit smaller longitudinal momentum shifts, whereas odd channels are weaker but carry larger momentum transfer. This ordering reverses at higher $I_n^{\mathrm{eff}}$.

\begin{figure}[t]
\centering
\includegraphics[width=\columnwidth]{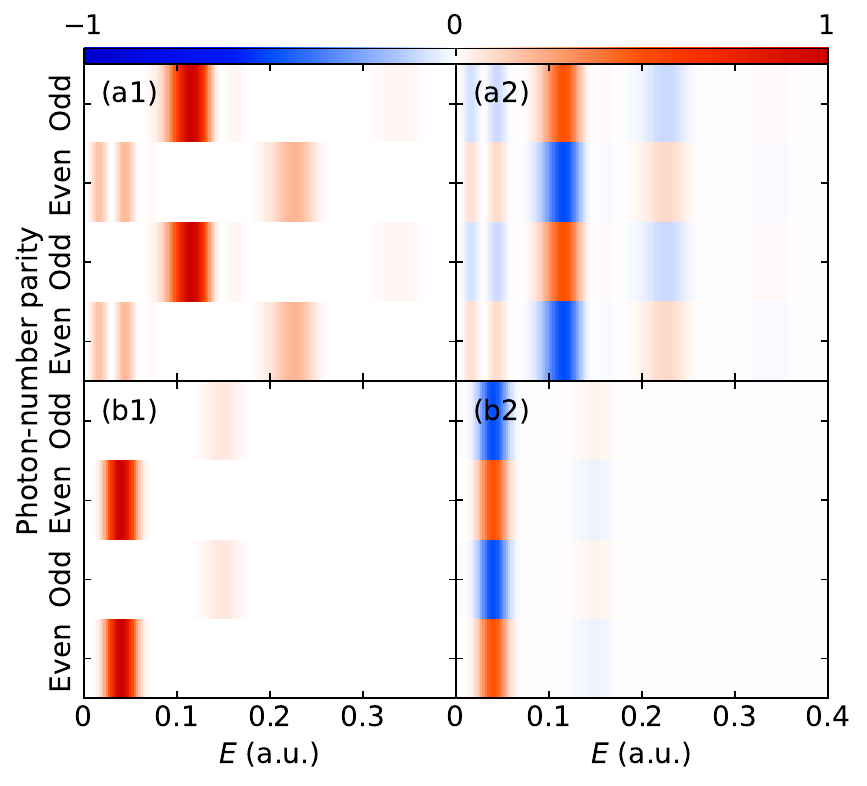}
\caption{
Saddle-point interference underlying the parity-dependent response described by Eq.~\eqref{eq:fock-interference-main}.
(a1) Conditional photoelectron spectrum $P_n(\mathbf p)$ and (a2) the corresponding interference contribution
$P_n(\mathbf p)-P_n^{\mathrm{inc}}(\mathbf p)$, both normalized to the maximum of $P_n(\mathbf p)$,
for higher-$n$ channels around $n\sim10^{13}$, corresponding to $I_n^{\mathrm{eff}}=9\times10^{13}\,\mathrm{W/cm^2}$.
(b1,b2) The same quantities as in (a1,a2), but for lower-$n$ channels around $n\sim10^{12}$, corresponding to $I_n^{\mathrm{eff}}=1\times10^{13}\,\mathrm{W/cm^2}$.
The remaining parameters are the same as in Fig.~\ref{fig:photon_resolved}.
}
\label{fig:photon_resolved_saddle}
\end{figure}

The parity dependence arises directly from the coherent addition of the two saddle contributions. A steepest-descent evaluation of Eq.~\eqref{eq:joint-main} about the two solutions in Eq.~\eqref{eq:fock-saddles-main} yields
\begin{equation}
P_n(\mathbf p)
=
P_n^{\mathrm{inc}}(\mathbf p)
+
2(-1)^n
\left|M_n^{(+)}(\mathbf p)M_n^{(-)}(\mathbf p)\right|
\cos\left[\Delta\varphi_n(\mathbf p)\right],
\label{eq:fock-interference-main}
\end{equation}
where $M_n^{(\pm)}(\mathbf p)$ denote the contributions of the two branches to the photon-number-projected ionization amplitude, $P_n^{\mathrm{inc}}(\mathbf p)=|M_n^{(+)}(\mathbf p)|^2+|M_n^{(-)}(\mathbf p)|^2$ is their incoherent sum, and the leading-order phase is
\begin{equation}
\Delta\varphi_n(\mathbf p)
=
\frac{\pi}{2\omega\Lambda}
\left(
\widetilde{\mathbf p}^{\,2}
+\kappa^2
+A_n^2
\right).
\label{eq:fock-phase-main}
\end{equation}
This dynamical phase accumulates over half an optical cycle according to the light-front quasienergy~\cite{Lin2022EnergyPeaks,He2022Subbarrier}, giving rise to double-slit interference in the time domain~\cite{PhysRevLett.95.040401}.
Interference extrema occur at $\Delta\varphi_n(\mathbf p)=N\pi$ with integer $N$, i.e.\ at
\begin{equation}
\frac{1}{\Lambda}
\left(
\frac{\widetilde{\mathbf p}^{\,2}}{2}
+I_p
+\frac{A_n^2}{2}
\right)
=
N\omega,
\label{eq:fock-ati-main}
\end{equation}
and the factor $(-1)^n$ renders these extrema constructive only for $N\equiv n\ (\mathrm{mod}\ 2)$. Neighboring photon-number channels therefore preferentially enhance complementary above-threshold-ionization combs.
Related even--odd structures observed in fully quantized one-dimensional simulations~\cite{Mao2025Benchmarking} provide a benchmark for this parity selection, while the nondipole momentum transfer addressed here lies outside the scope of the one-dimensional dipole model employed there.

We next examine how this parity-selective interference is converted into the observed longitudinal momentum response. Figure~\ref{fig:photon_resolved_saddle} presents the spectra and the corresponding interference contributions obtained from Eq.~\eqref{eq:fock-interference-main}, illustrating how the interference-induced spectral redistribution generates the parity-dependent momentum response. The subpeaks visible in Figs.~\ref{fig:photon_resolved_saddle}(a1) and \ref{fig:photon_resolved_saddle}(a2) arise from the time-dependent ponderomotive shift induced by the pulse envelope~\cite{PhysRevA.73.011401}. For lower-$n$ channels, the interference term predominantly suppresses the low-energy spectral weight in odd channels while enhancing it in even ones. Consequently, the odd channels acquire a higher characteristic photoelectron energy, a larger longitudinal momentum transfer, and a broader longitudinal momentum distribution, consistent with the established connection between continuum energy and nondipole radiation pressure~\cite{Smeenk2011PhotonMomentum,Reiss2013RelativisticEffects,Chelkowski2014MomentumSharing,He2017BeyondDipole,supp}. For higher-$n$ channels, the ponderomotive shift and successive channel closings~\cite{milovsevic2006above} reorganize the parity-selected interference structure across the photoelectron-energy spectrum. This redistribution shifts the regions of constructive enhancement and destructive suppression, reversing the even--odd ordering of the characteristic photoelectron energy and longitudinal momentum transfer relative to the lower-$n$ regime. Viewed together with the photon-traced result, this analysis shows that the $I^{2/3}$ radiation-pressure law in Eq.~\eqref{eq:momentum-scaling-main} is the phase-insensitive remnant of the underlying two-branch coherent structure.

The photon-traced radiation-pressure anomaly in Eq.~\eqref{eq:momentum-scaling-main} can be tested directly by measuring the photoelectron momentum distribution along the laser-propagation direction.
Recent experiments have generated BSV pulses with mean peak intensities up to $1.6\times10^{14}\,\mathrm{W/cm^2}$~\cite{Liu2026BSVHolography}, reaching the regime in which nondipole longitudinal momentum shifts are experimentally resolvable~\cite{Hartung2019Magnetic}. By contrast, accessing the photon-number-conditioned parity response is more demanding. A possible route is to weakly tap the outgoing BSV field and combine electron--photon coincidence detection with parity-sensitive heralding based on one- and two-photon subtraction, techniques already demonstrated for squeezed vacuum~\cite{Takahashi2008PhotonSubtraction,Gerrits2010NumberResolvedSubtraction}. Shot-resolved photon-number conditioning has also been implemented in strong-field photoelectron and high-harmonic measurements~\cite{Heimerl2025QuantumNeedle,Tsatrafyllis2017IRPhotonStatistics}, while related conditional protocols have been proposed for generating nonclassical optical states through strong-field interactions~\cite{Lewenstein2021OpticalCats}. These developments suggest possible routes to probing the parity-dependent response, although high detection efficiency and stringent loss control would be required to preserve the even--odd contrast.

In summary, we have shown that the radiation pressure exerted by BSV on photoelectrons follows an anomalous $\langle p_z\rangle\propto I^{2/3}$ law arising from nonlinear tunneling that selects two field configurations of equal magnitude and opposite sign from the broad BSV amplitude distribution. 
When the photon state is traced out, the relative-phase information between the two branches is lost, while their common magnitude sets the radiation-pressure scale. When the photon number is resolved, the two branches interfere as field pathways separated by half an optical cycle, producing parity-dependent modulations of the momentum transfer. 
Radiation pressure thus probes not only the amplitude statistics of intense quantum light but also the coherence among its field configurations, and this sensitivity extends beyond BSV to other broadband quantum states of light.

\begin{acknowledgments}
This work was supported by the National Natural Science Foundation of China (NSFC) under Grant Nos.~12574377 and 12450405. The computations were performed on the Siyuan-1 cluster supported by the Center for High Performance Computing at Shanghai Jiao Tong University. P.-L. H. acknowledges support from the Pujiang Program of the Shanghai Baiyulan Talent Plan (Grant No.~24PJA046), the Xiaomi Young Scholar Program, the Shanghai Jiao Tong University 2030 Initiative, and the Yangyang Development Fund. X. M. acknowledges support from the Postdoctoral Fellowship Program of CPSF under Grant No.~GZB20260818, the China Postdoctoral Science Foundation under Grant No.~2026M793723, and the Shanghai Super Postdoctoral Fellowship Program.
\end{acknowledgments}

\bibliography{references}

\clearpage
\section*{End Matter}
In this End Matter, we develop two complementary saddle-point constructions underlying the main results. For photon-traced observables, the saddle analysis is performed at the probability level and determines the field-amplitude configurations responsible for the anomalous $I^{2/3}$ radiation-pressure scaling. For photon-number-resolved observables, the analysis is performed at the amplitude level and yields a coherent pair of field saddles whose interference produces the parity-dependent response. We treat these two cases in turn.

\subsection{Photon-traced saddle structure}
\label{sec:end-Q-saddles}

For the circularly polarized field considered in the main text, the exponent governing the photon-traced probability in Eq.~\eqref{eq:field-average-main} is
\begin{equation}
\Phi_Q
=
-\frac{A_\alpha^2}{2A_{\mathrm{rms}}^2}
+
\frac{i}{2\Lambda}
\int_{\eta_2}^{\eta_1}d\tau\,
\left\{
\left[\widetilde{\mathbf p}+\mathbf A(\tau;A_\alpha)\right]^2+\kappa^2
\right\}.
\label{eq:end-Q-exponent}
\end{equation}
Requiring $\Phi_Q$ to be stationary with respect to the light-front ionization times $\eta_1$, $\eta_2$, and the field amplitude $A_\alpha$ yields the complete set of saddle-point equations. The field-amplitude saddle-point equation is
\begin{equation}
\frac{A_s^2}{A_{\mathrm{rms}}^2}
=
\frac{i}{\Lambda}
\int_{\eta_{2,s}}^{\eta_{1,s}}d\tau\,
\left[
\widetilde{\mathbf p}
+\mathbf A(\tau;A_s)
\right]\cdot
\mathbf A(\tau;A_s),
\label{eq:end-Q-saddle-system}
\end{equation}
and the ionization-time saddle-point equations are
\begin{align}
&\left[\widetilde{\mathbf p}+\mathbf A(\eta_{j,s};A_s)\right]^2+\kappa^2=0,
\qquad j=1,2.
\label{eq:end-Q-saddle-system-times}
\end{align}

For real photoelectron momenta, the full saddle-point system is invariant under the conjugation transformation
\begin{equation}
(A_s,\eta_{1,s},\eta_{2,s})
\longmapsto
(A_s^*,\eta_{2,s}^*,\eta_{1,s}^*).
\label{eq:end-Q-conjugation}
\end{equation}
The saddle solutions therefore consist of self-conjugate solutions and genuinely complex-conjugate pairs. In the parameter regime considered in the main text, the dominant solutions belong to the self-conjugate sector,
\begin{equation}
A_s\in\mathbb R,\qquad
\eta_{1,s}=\eta_r+i\eta_i,\qquad
\eta_{2,s}=\eta_r-i\eta_i.
\label{eq:end-Q-self-conjugate}
\end{equation}
This self-conjugate system can be reduced analytically to a compact form. We begin with the ionization-time saddle-point equation. 
We introduce the real and imaginary saddle phases
\[\phi_r=\omega\eta_r-\theta/2,\qquad
\phi_i=\omega\eta_i.\]
The transformation
$(A_s,\phi_r)\rightarrow(-A_s,\phi_r+\pi)$
leaves the vector potential invariant. We therefore choose the representative branch with
$A_s>0$ and $\phi_i>0$.

The imaginary part of Eq.~\eqref{eq:end-Q-saddle-system-times} requires
\begin{equation}
\left[\mathbf p_\perp+\operatorname{Re}\bm A_\perp\right]\cdot
\operatorname{Im}\mathbf A_\perp=0,
\end{equation}
where $\mathbf p_\perp=(p_x,p_y)$ denotes the component of $\widetilde{\mathbf p}$ in the laser-polarization plane.
We therefore parameterize
\begin{align}
&\mathbf p_\perp+\operatorname{Re}\mathbf A_\perp(\eta_s;A_s) \nonumber \\
=&
 v_\perp
\frac{
-\operatorname{Im}A_y(\eta_s;A_s)\,\mathbf e_x
+\operatorname{Im}A_x(\eta_s;A_s)\,\mathbf e_y
}{
\left|\operatorname{Im}\mathbf A_\perp(\eta_s;A_s)\right|
},
\end{align}
where $v_\perp$ is a real-valued parameter; this parametrization satisfies the imaginary-part condition identically. Using the explicit form of the circularly polarized vector potential then gives
\begin{equation}
\begin{aligned}
p_x&=-\left(A_s\cosh\phi_i+v_\perp\right)\sin\phi_r,\\
p_y&=\left(A_s\cosh\phi_i+v_\perp\right)\cos\phi_r.
\end{aligned}
\label{eq:end-Q-momentum-components}
\end{equation}
The real part of the ionization-time saddle-point equation then reduces to
\begin{equation}
A_s\sinh\phi_i=\widetilde\kappa,\qquad
\widetilde\kappa=\sqrt{\kappa^2+\widetilde p_z^2+v_\perp^2}.
\label{eq:end-Q-circular-time}
\end{equation}
Combining
Eqs.~\eqref{eq:end-Q-momentum-components}
and~\eqref{eq:end-Q-circular-time}
with the field-amplitude saddle-point equation [Eq.~\eqref{eq:end-Q-saddle-system}] gives
\begin{equation}
\frac{\omega\Lambda}{A_{\mathrm{rms}}^2}
=
\sinh(2\phi_i)-2\phi_i
+
2\frac{v_\perp}{\widetilde\kappa}\sinh^2\phi_i .
\label{eq:end-Q-circular-field}
\end{equation}
Equations~\eqref{eq:end-Q-momentum-components}--\eqref{eq:end-Q-circular-field} constitute the reduced self-conjugate saddle system used in the numerical calculations and retain the full nonadiabatic dependence, allow for a finite transverse exit velocity, and include the leading light-front correction.

In the adiabatic tunneling limit, $\phi_i\ll1$, the solution for $v_\perp=0$, including the equivalent opposite-sign branch, becomes
\begin{equation}
A_s\simeq
\pm\sqrt{\kappa^2+\widetilde p_z^2}
\left(
\frac{4A_{\mathrm{rms}}^2}{3\omega\Lambda}
\right)^{1/3},
\qquad
A_s^2\propto I^{2/3}.
\label{eq:end-Q-adiabatic-As}
\end{equation}
This result directly reveals the origin of the anomalous intensity scaling.

\begin{figure*}[t]
\centering
\includegraphics[width=\textwidth]{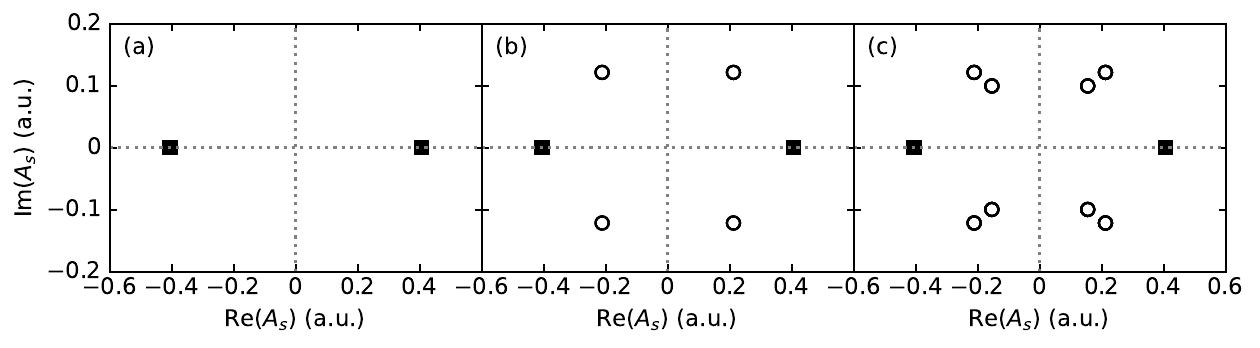}
\caption{
Field saddles in the complex-$A_s$ plane for the representative momentum
$(p_x,p_y,p_z)=(A_{\mathrm{rms}},0,A_{\mathrm{rms}}^2/2c+I_p/3c)$,
with $A_{\mathrm{rms}}=0.3$ a.u.
Panels (a)--(c) show saddle pairs satisfying
$\left|\operatorname{Re}\eta_{1,s}-\operatorname{Re}\eta_{2,s}\right|<mT$,
with $m=1$, $2$, and $3$, respectively.
Squares denote self-conjugate saddles, and circles denote genuinely complex saddles.
The remaining parameters are the same as in Fig.~\ref{fig:pzshift}.
}
\label{fig:end-saddles}
\end{figure*}

Beyond the dominant self-conjugate solutions, the saddle equations also admit genuinely complex saddles occurring in conjugate pairs. Figure~\ref{fig:end-saddles} shows the locations of the self-conjugate and genuinely complex saddles in the complex-$A_s$ plane as the allowed separation between the real parts of the two saddle times is progressively increased. Panels (a)--(c) include saddle pairs satisfying
\[
\left|\operatorname{Re}\eta_{1,s}
-\operatorname{Re}\eta_{2,s}\right|
<mT,
\qquad
m=1,2,3,
\]
respectively. Simultaneous translations of both saddle times by the same integer multiple of $T$ generate optical-cycle replicas and are not counted as distinct solutions.

For the representative parameters used here, each increase of the cutoff from $mT$ to $(m+1)T$ over the range shown adds four saddle solutions, corresponding to two additional complex-conjugate pairs. These newly included solutions lie on intercycle branches for which the real parts of the two saddle times differ by one or more optical periods. They generally have genuinely complex field amplitudes, $\operatorname{Im}A_s\neq0$.

The marker positions in Fig.~\ref{fig:end-saddles} encode only the saddle locations, not their weights. A separate evaluation of the exponential weights $\exp[\operatorname{Re}\Phi_Q]$, normalized to that of the dominant self-conjugate saddle, shows rapid suppression with increasing temporal separation. For the representative parameters used here, the additional intercycle saddles that first appear in panel (b) carry relative weights of order $10^{-3}$, whereas those appearing only in panel (c) carry relative weights of order $10^{-5}$, corresponding to suppressions of roughly three and five orders of magnitude, respectively.
These small subleading contributions partly account for the residual deviations from the near-universal $\sigma_{p_z}$--$\langle p_z\rangle$ relation shown in Fig.~\ref{fig:scaling}.

\subsection{Photon-number-resolved saddle structure}
\label{sec:end-path-integral}

When the final photon number is resolved, the coherent-state amplitudes must be summed before taking the modulus squared. 
Introducing \(\beta=\alpha e^{-i\theta/2}\) and \(\bar\beta=\bar\alpha e^{i\theta/2}\), we write the amplitude-level exponent as
\begin{equation}
\Phi_n(\beta,\bar\beta;\eta)
=
-iS(\beta,\bar\beta;\eta)
+n\ln\beta
-\beta\bar\beta
-\frac{1}{2}\bar\beta^2\tanh r,
\label{eq:end-path-exponent}
\end{equation}
where the circularly polarized vector potential is
\begin{equation}
\mathbf A_\beta(\eta)
=
\epsilon_V
\left[
\beta e^{-i\phi(\eta)}
\left(\mathbf e_x+i\mathbf e_y\right)
+
\bar\beta e^{i\phi(\eta)}
\left(\mathbf e_x-i\mathbf e_y\right)
\right].
\end{equation}
Defining
\(\boldsymbol{\Pi}_\perp(\tau)
=\mathbf p_\perp+\mathbf A_\beta(\tau)\), we write the semiclassical action as
\begin{equation}
S(\beta,\bar\beta;\eta)
=
\int_\eta^\infty
\frac{d\tau}{2\Lambda}
\left[
\boldsymbol{\Pi}_\perp^{\,2}(\tau)
+\widetilde p_z^2+\kappa^2
\right].
\end{equation}

Treating \(\beta\) and \(\bar\beta\) as independent variables and requiring stationarity with respect to \(\beta\), \(\bar\beta\), and \(\eta\) yields
\begin{equation}
\begin{aligned}
-i\mathcal J_-+\frac{n}{\beta}-\bar\beta&=0,\\
-i\mathcal J_+-\beta-\bar\beta\tanh r&=0,
\\
\boldsymbol{\Pi}_\perp^{\,2}(\eta_s)
+\widetilde p_z^2+\kappa^2&=0,
\label{eq:end-path-saddles}
\end{aligned}
\end{equation}
where $\mathcal J_\pm
=
\epsilon_V
\int_{\eta_s}^{\infty}
\frac{d\tau}{\Lambda}
\left[
\Pi_x(\tau)\mp i\Pi_y(\tau)
\right]
e^{\pm i\phi(\tau)}$.

In the bright-field, large-$n$ regime relevant here, $\tanh r\simeq1$, and for the parameters considered $\epsilon_V\sim10^{-8}$~a.u. Expanding the saddle equations to first order in the electron-induced displacement gives
\begin{align}
\beta_s^{(\pm)}
&\simeq
\pm i\sqrt n
-\frac{i}{2}
\left(
\mathcal J_+-\mathcal J_-
\right),
\nonumber\\
\bar\beta_s^{(\pm)}
&\simeq
\mp i\sqrt n
-\frac{i}{2}
\left(
\mathcal J_++\mathcal J_-
\right).
\label{eq:end-path-saddle-pair}
\end{align}
At leading order,
\(\beta_s^{(\pm)}=\pm i\sqrt n\) and
\(\bar\beta_s^{(\pm)}=\mp i\sqrt n\). 
These leading-order saddles generate the two opposite fields
\begin{equation}
\mathbf A_\perp^{(\pm)}(\eta)
=
\pm A_n
\left[
\sin\phi(\eta)\,\mathbf e_x
-\cos\phi(\eta)\,\mathbf e_y
\right].
\label{eq:end-path-opposite-fields}
\end{equation}

For a monochromatic circular field, the two opposite-field branches are related by a half-cycle shift,
\begin{equation}
\phi_s^{(-)}
=
\phi_s^{(+)}+\pi
\qquad
(\mathrm{mod}\ 2\pi).
\label{eq:end-path-half-cycle}
\end{equation}
Evaluating the exponent at the saddle pair in Eq.~\eqref{eq:end-path-saddle-pair} and using the half-cycle relation in Eq.~\eqref{eq:end-path-half-cycle} yields the interference structure given by Eqs.~\eqref{eq:fock-interference-main} and \eqref{eq:fock-phase-main} of the main text. The term \(n\ln\beta\) contributes a relative phase of \(n\pi\), producing the factor \((-1)^n\), while the difference between the two Volkov actions gives the continuous dynamical phase in Eq.~\eqref{eq:fock-phase-main}.

\end{document}